\documentclass[prd, twocolumn, nofootinbib]{revtex4}

\usepackage{epsfig} 

\newcommand{\beq}{\begin{equation}}
\newcommand{\eeq}{\end{equation}}
\newcommand{\beqa}{\begin{eqnarray}}
\newcommand{\eeqa}{\end{eqnarray}}
\newcommand{\om}{\Omega_m}

\newcommand{\eps}{\epsilon} 
\newcommand{\vp}{V_{,\phi}}
\newcommand{\vpp}{V_{,\phi\phi}}
\newcommand{\rhos}{\rho_\star}

\def\fun#1#2{\lower3.6pt\vbox{\baselineskip0pt\lineskip.9pt
  \ialign{$\mathsurround=0pt#1\hfil##\hfil$\crcr#2\crcr\sim\crcr}}}

\begin{document} 

\title{The Paths of Quintessence} 
\author{Eric V.\ Linder} 
\affiliation{Berkeley Lab, University of California, Berkeley, CA 94720} 

\begin{abstract} 
The structure of the dark energy equation of state phase plane 
holds important information on the nature of the physics. 
We explain the bounds of the freezing and thawing models of 
scalar field dark energy in terms of the tension between 
the steepness of the potential vs.\ 
the Hubble drag.  Additionally, we extend the phase plane structure 
to modified gravity theories, examine trajectories 
of models with certain properties, and categorize 
regions in terms of scalar field hierarchical parameters, showing 
that dark energy is generically not a slow roll phenomenon. 

\end{abstract} 


\date{\today}

\maketitle

\section{Introduction} \label{sec.intro}

The discovery of the acceleration of the cosmic expansion has thrown 
physics and astronomy research into a ferment of activity, from a 
search for fundamental theories to investigation of predictions 
relating models and the cosmological dynamics, to development of 
astrophysical surveys yielding improved measurements.  

The acceleration, or more generally the expansion history 
of the scale factor evolution with time, $a(t)$, can be equivalently 
treated by a dark energy pressure to density, or equation of state, 
ratio $w(a)$ \cite{linjen}.  One model, Einstein's cosmological constant, 
predicts $w=-1$ at all times, but generically the dark energy phenomenon 
has dynamics, a time varying $w(a)$.  

It is important to note that the 
current epoch of accelerated expansion is very different from the early 
epoch of inflation.  We know a priori that dark energy does not completely 
dominate the universe now and we do not know a priori that dark energy 
obeys a slow roll approximation (in fact we will see it is unlikely to). 

In these senses, dark energy is a more challenging phenomenon than 
inflation.  We are faced with three ``Goldilocks'' problems: 1) dark 
energy is dynamically important, but not fully dominant, with $\sim75\%$ 
of the total energy density today, 2) the universe is accelerating, as 
from a component with $w\lesssim-0.8$, but the field responsible may not 
be slowly rolling (unless fine tuned) as it would if nearly completely 
potential dominated, and 3) if it's not the cosmological 
constant, what happened to the cosmological constant? 

To discover whether the physics 
is the cosmological constant, and to distinguish between alternate 
theories, requires measurement of the possible dynamics.  

Caldwell \& Linder \cite{caldlin} (Paper 1) investigated the dynamics in 
the phase plane of $w$-$w'$, where $w'=dw/d\ln a$, 
for canonical scalar field dark energy, or quintessence, finding that this 
reveals important clues to the nature of the new physics.  In such a plane, 
the time or scale factor variable is a parameter along the paths of 
dynamics.  They found distinct structure, categorizing fields into those 
that at early times are locked by Hubble friction into a cosmological 
constant like state, and then move away from this (thaw) as the dark 
energy dominates, and those that initially roll due to the steepness 
of the potential but later approach the cosmological constant (freeze). 
In this article some of these results are put on a firmer 
footing, examined in greater detail, and extended to models beyond 
canonical scalar fields, including modified gravity theories.

\section{General Dynamics} \label{sec:scadyn} 

We begin with a brief, explicit derivation of the key dynamics 
equation, Eq.~(\ref{eq:wp}).  
A scalar field $\phi$ possesses both potential energy 
$V(\phi)$ and kinetic energy $(1/2)\dot\phi^2$.  The Lagrangian 
density is of the form 
\beq 
{\mathcal L}=(1/2)g_{\mu\nu}\partial^\mu\phi\partial^\nu\phi-V(\phi), 
\eeq 
for a canonical, minimally coupled scalar field in a metric $g_{\mu\nu}$.  
The equation of motion for the field, the Klein-Gordon equation, 
\beq 
\ddot\phi+3H\dot\phi+V_{,\phi}=0, \label{eq:kg} 
\eeq 
where $H=\dot a/a$ is the Hubble parameter, 
follows from functional variation of the Lagrangian.  (Spatial 
inhomogeneities should be negligible on subhorizon scales, but also 
see \S\ref{sec:inhom}.)  The energy-momentum 
tensor is generated through Noether's theorem and one can identify the 
energy density and pressure: 
\beq 
\rho=(1/2)\dot\phi^2+V(\phi)\quad;\quad p=(1/2)\dot\phi^2-V(\phi). 
\eeq 
It will be convenient to invert these and write the potential and 
kinetic energies in terms of $\rho$ and $w$: 
\beq 
V=(1/2)(1-w)\rho\quad;\quad (1/2)\dot\phi^2=(1/2)(1+w)\rho. \label{eq:vk} 
\eeq 
Note that $w$ and $\rho$ are both functions of time, or scale factor. 

To obtain an equation for the  variation $w'$, 
take the derivative of $V$ with respect to $\phi$, 
\beq 
\vp=\dot V/\dot\phi=(1/2)[(1-w)\dot\rho-\rho\dot w]/[(1+w)\rho]^{1/2}. 
\label{eq:vdot} 
\eeq 
Employing the continuity equation $\dot\rho=-3H\rho(1+w)$ and 
$H=d\ln a/dt$, one obtains 
\beq 
w'=-3(1-w^2)-(1-w)(1+w)^{1/2}\sqrt{(3/8\pi) \Omega_\phi(a)}\frac{M_P \vp}{V}, 
\label{eq:wp} 
\eeq 
where $\Omega_\phi(a)=8\pi\rho/(3H^2M_P^2)$ is the dimensionless dark energy 
density, and $M_P$ is the Planck mass. 

\subsection{Distinguishing $\Lambda$ \label{sec:barrier}} 

We see that $w'$ is related to the nearness of the equation of state 
ratio to the cosmological constant value, i.e.\ $1+w$, and the inverse 
of the characteristic field scale of the potential, $\vp/V$ 
(sometimes phrased as a slow roll parameter, $M_P|\vp/V|\ll1$).  
If $w$ is readily distinguished from $-1$, then we know the dark energy 
is not a cosmological constant, regardless of the value of the time 
variation $w'$.  More difficult is the case where $\eps=1+w$ is a small 
quantity.  Then it will be quite important to determine whether $w'$ is 
zero or not.  Eq.\ (\ref{eq:wp}) guides us in following the dynamics 
in the $w$-$w'$ phase space.  

\begin{figure}[!hbt]
\begin{center} 
\psfig{file=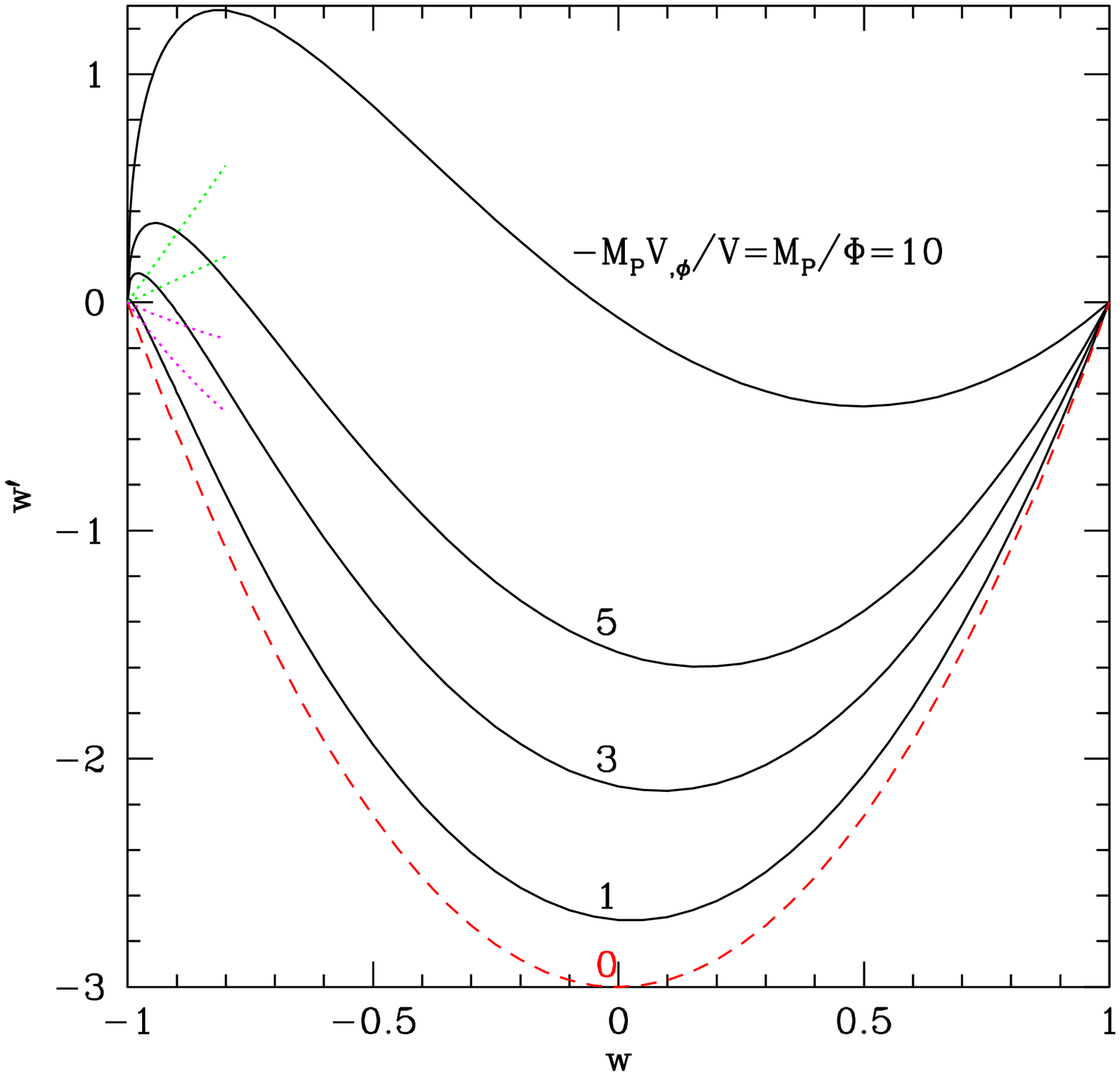,width=3.4in} 
\psfig{file=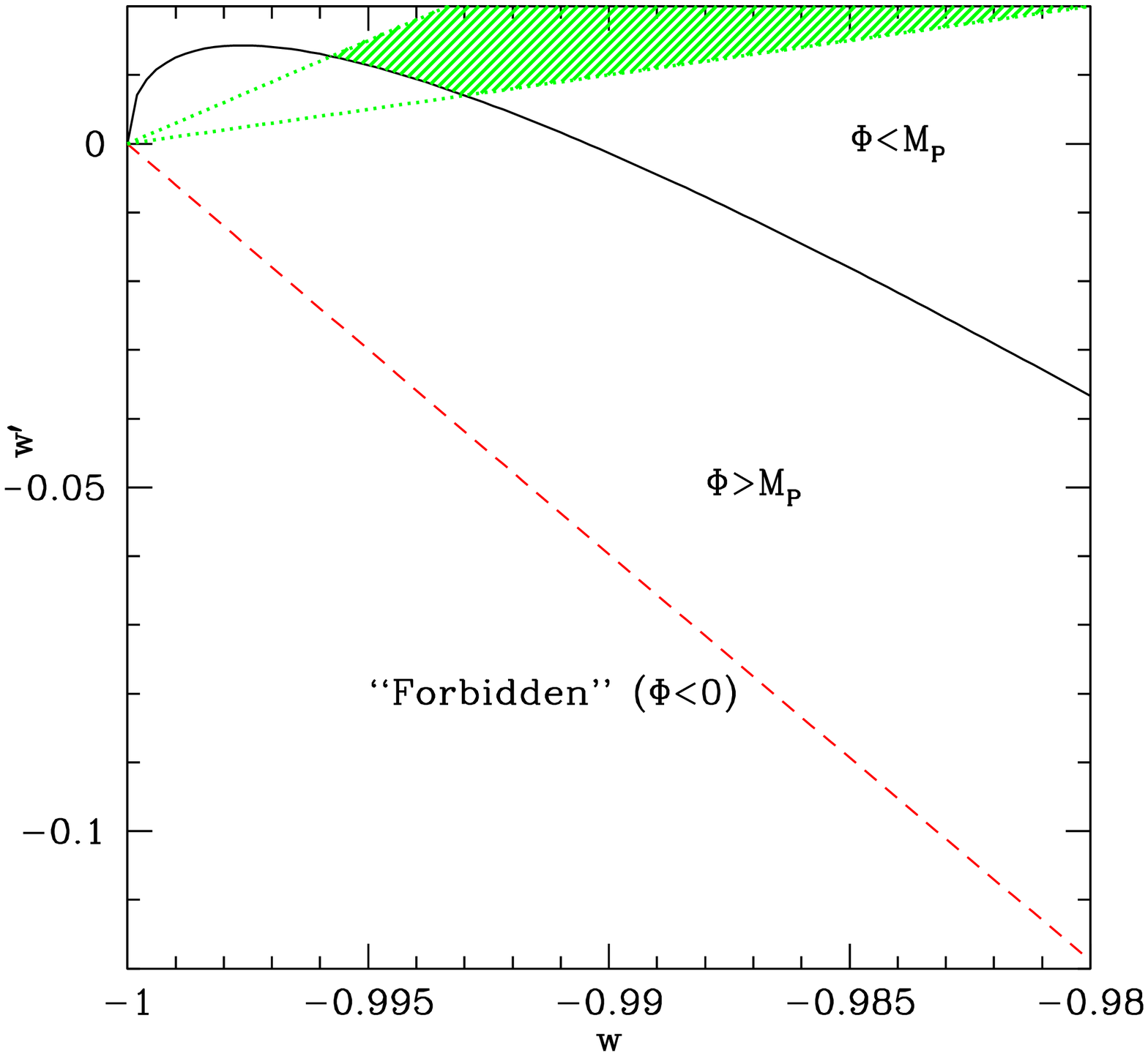,width=3.4in} 
\caption{Contours of the inverse of the characteristic field scale 
$\Phi=-V/\vp$ are plotted in full and zoomed versions, for present 
matter density $\om=0.28$.  Note that only 
a slender crescent of the phase plane obeys the conventional slow roll 
criterion $M_P/\Phi\ll1$.  The dashed red curve shows the null 
line $\vp=0$.  Dotted lines show the $w\approx-1$ limits of the 
thawing region (upper pair: $w'=3(1+w), (1+w)$) and freezing 
region (lower pair: here $w'=w(1+w), 3w(1+w)$; not shown in 
bottom panel). If physics limits the characteristic field scale to be 
less than the Planck mass, then a barrier forms (second panel) 
around the cosmological constant, allowing only models in the shaded  
part of the thawing region. 
}
\label{fig:wwpscale} 
\end{center} 
\end{figure}

The first point to notice is that the reciprocal of the characteristic 
field scale is not generically a small parameter useful for a ``slow roll'' 
approximation.  
Figure \ref{fig:wwpscale} shows curves of constant field scale 
\beq 
\Phi=V/(-\vp), 
\eeq 
in the $w$-$w'$ plane.  Only a tiny sliver of the phase space, plus a 
small hump, satisfies the slow roll approximation; 
unless one is exceedingly close to the cosmological constant behavior 
there is substantial dynamics in the field. 

When $1+w\ll1$, the second term in Eq.~(\ref{eq:wp}) 
dominates and $w'>0$ (creating the ``humps''), while for less 
negative $w$ the first term can 
dominate and drive $w'<0$.  This driving occurs closer to $w=-1$ for larger 
$\Phi$.  Within the large-$\Phi$ hump, the dark energy looks similar to a 
cosmological constant.  Suppose one conjectures some physics limiting 
the size of the field scale, equivalently leading to avoidance of 
flatness in the potential.  An upper bound on $\Phi$ in the scalar 
field behavior 
would impose a barrier around the cosmological constant $\Lambda$, saying 
that the scalar field dynamics must be distinguishable from $\Lambda$, 
if it is not $\Lambda$. 

The second panel of Fig.~\ref{fig:wwpscale} zooms in to illustrate 
this barrier for 
$\Phi<M_P$, which rules out any freezing field, and any thawing field 
with $1+w<0.004$.  So any scalar field theory with field scales barred 
from being transPlanckian are distinguishable from $\Lambda$ at this 
precision.  If the restriction uses, say, $\Phi<M_P/\sqrt{8\pi}$ (i.e.\ 
the Planck mass is defined in terms of Newton's constant as 
$G_N=(8\pi M_P^2)^{-1}$ rather than $G_N=M_P^{-2}$ as 
above) then the limit becomes $1+w>0.1$.  

The physical origin for the conjectured limit on the characteristic 
field scale is not clear, but the implications are important enough 
to consider the possibility. 
Since the field scale is related to the inverse of the flatness of 
the potential, then physics that perturbs a flat direction in the potential 
would give this effect.  In some supersymmetric models, loop 
corrections generate a logarithmic tilt $V\sim\phi^n\ln(\phi/\mu)$ 
\cite{witten,damourmukhanov}.  This would give $\Phi\approx\phi/n$ (or 
$\phi\,\ln(\phi/\mu)$ for $n=0$), 
and restricting $\phi<M_P$ (for the effective field theory to be valid), 
provides the limit $\Phi\lesssim{\cal O}(M_P)$.  However, the generic 
breakdown of slow roll is independent of any $\Phi$ upper limit conjecture 
and we do not consider the latter further. 
Hierarchical parameters 
to replace slow roll are discussed in \S\ref{sec:slow}.

\subsection{Driving and dragging \label{sec:kglines}}

Returning to the Klein-Gordon equation, we can understand behavior 
in the $w-w'$ phase space by first a general and then a specific 
analysis of the terms.  Writing Eq.~(\ref{eq:kg}) as 
$\ddot\phi+3H\dot\phi=-\vp$, we can require $-\vp\ge0$, reflecting 
that the field rolls down the potential to large $\phi$.  Using 
Eq.~(\ref{eq:vk}) for $\dot\phi$, we write this condition in 
terms of $w$, $w'$ as $w'\ge-3(1-w^2)$.  The boundary defines the 
null line $\vp=0$ discussed further below.  Similarly, writing the 
Klein-Gordon equation as $\ddot\phi+\vp=-3H\dot\phi$ and again using 
that the field rolls to larger values with time implies $-3H\dot\phi\le0$. 
This reduces to the condition $w\ge-1$.  

If we flip the direction of 
the field motion and potential slope, i.e.\ the field rolls down the 
potential to smaller $\phi$, then these conditions remain. 
(I.e.\ $\dot V=\dot\phi\vp$ is still negative, and so the transition 
from Eq.~(\ref{eq:vdot}) to Eq.~(\ref{eq:wp}) flips the sign of the 
second term on the right hand side of Eq.~(\ref{eq:wp}), canceling 
the reversed sign of $\vp$.)  
However, if we make the field move up the potential then we have 
the relations $w'\le-3(1-w^2)$ and $w<-1$, in the phantom region, 
i.e.\ the boundary lines just continue smoothly through $w=-1$. 

Finally, we can move $\ddot\phi$ to the right hand side to obtain 
$3H\dot\phi+\vp=-\ddot\phi$.  This divides the $w-w'$ plane into 
regions where the field is accelerating or decelerating, with the 
boundary being the coasting behavior $\ddot\phi=0$.  This line 
corresponds to the condition $w'=3(1+w)^2$.  Larger values of $w'$ 
arise from a field accelerating down the potential, while smaller values 
come from a field decelerating (this motion of the field should not 
be confused with the accelerating expansion of the universe, which 
can hold for either region).  These three boundaries -- the null 
line $w'=-3(1-w^2)$, coasting line $w'=3(1+w)^2$, and phantom line 
$w=-1$ -- define the broad characteristics of the phase plane. 

To investigate the dynamics further, we must examine the dominance 
of the different terms in the equation of motion (\ref{eq:kg}).  
The first term is the acceleration of the field, the second 
term a friction term, or Hubble drag, due to the expansion of the 
universe, and the third is a driving term due to the steepness 
of the potential.  We define 
\beqa 
X&=&\frac{\ddot\phi}{H\dot\phi} \\ 
Y&=&\frac{\ddot\phi}{\vp}\ . 
\eeqa 

\begin{figure}[!hbt]
\begin{center}
\psfig{file=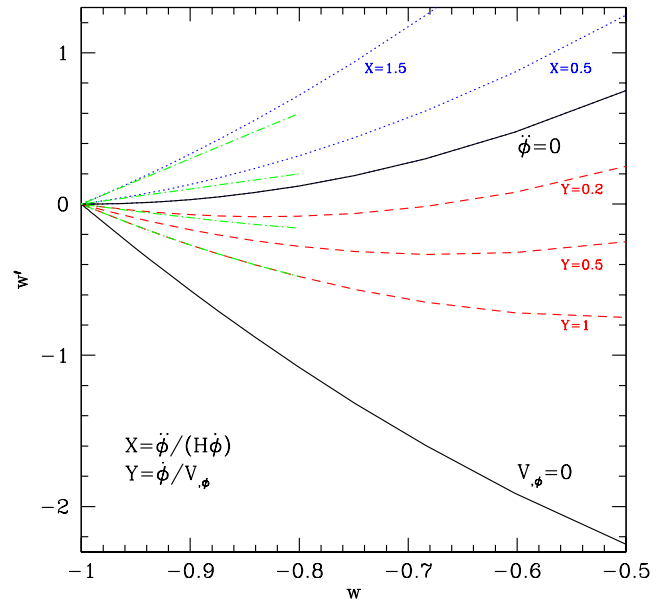,width=3.4in}
\caption{The physics behind the dynamics becomes more apparent when 
looking at the ratios of the terms in the Klein-Gordon equation of 
motion, e.g.\ field acceleration vs.\ friction vs.\ potential driving. 
The key general features of the phase plane are the null line $\vp=0$, 
the coasting line $\ddot\phi=0$, and the phantom line $w=-1$.  Thawing 
(with $w'>0$) and freezing (with $w'<0$) regions are bounded by 
green, dot-dash curve pairs and are defined through the physical dynamics. 
}
\label{fig:kgterms}
\end{center}
\end{figure}

Figure~\ref{fig:kgterms} shows curves of constant $X$, $Y$ in the 
$w-w'$ plane.  Note that $X=Y=0$ corresponds to an epoch of coasting 
in the scalar field dynamics, $\ddot\phi=0$, as discussed above.  
This is nongeneric, 
as the field would need to be finely tuned to neither accelerate 
due to the slope of the potential nor decelerate due to the Hubble drag, 
but be perfectly balanced. 
Indeed, the dynamics of scalar fields in Paper 1 avoid this 
region, causing the split into the distinct thawing and freezing regions, 
respectively above and below this line. 

In the accelerating field region, the friction term is the major 
determinant of behavior initially, as the field evolves away from a 
frozen (cosmological constant-like) state in the matter dominated epoch. 
The upper boundary of the thawing region is given by $X=3/2$, where 
this value follows directly from the exponent of the expansion history, 
$t\sim a^{3/2}$ for matter domination.  Thus fields that thaw during 
matter domination begin to move along the $X=3/2$ line (see discussion, 
and Figure 2, in Paper 1).  

We can translate any acceleration to friction ratio $X$ to a $w-w'$ 
behavior through 
\beqa 
w'&=&2X(1+w)+3(1+w)^2=(1+w)(3+2X+3w) \nonumber \\ 
&\approx& 2X(1+w), 
\eeqa 
where the last approximation is for $1+w\ll1$.  Note that the linear 
boundaries used in Paper 1 were (good) approximations to the general 
parabolic behavior.  The value $X=0$ gives the coasting line and $X=-3$ 
gives the null line. 

Thus, the upper thawing boundary $X=3/2$ corresponds to $w'\approx 3(1+w)$ 
for $1+w\ll1$.  If the rolling field then enters a region where the 
potential slope is shallower (as usually happens), then the field will 
accelerate less and curve toward the $\ddot\phi=0$ line.  Since today the 
field cannot have rolled so far that $\Omega_{\rm de}>0.8$, the 
dynamical track remains within the thawing region $1+w<w'<3(1+w)$, 
i.e.\ $1/2<X<3/2$.  
For potentials that steepen as the field rolls down, e.g.\ PNGB models 
with the field starting near the top of the potential, the tracks instead 
lie above the $X=3/2$ or $w'=3(1+w)$ line.  The PNGB potential also steepens 
more rapidly for small symmetry scales $f$, and indeed, as mentioned in 
Linder \cite{osc}, PNGB models roughly follow $w'=F(1+w)$, where $F$ 
is proportional to the inverse of $f$.  (Also see \cite{kalopersorbo} 
for discussion of PNGB models, fine tuning, and slow roll.)  

In the decelerating field region, the steepness of the potential impacts 
the freezing.  As the potential becomes shallower, the friction is more 
effective.  In the limit of a flat potential, one obtains the dynamics 
track given by the null curve, $\vp=0$, in Fig.~\ref{fig:kgterms}.  As 
given in Paper 1, this corresponds to $w'=-3(1-w^2)$, and 
the skating model of \cite{lincurv,liddleskate}.  
In terms of the friction and driving terms, $X=-3$ and $Y=\infty$. 
The lower boundary of the freezing region lies along the $Y=1$ ($X=-3/2$) 
line, equivalent to $w'=3w(1+w)$.  (See \S\ref{sec:track} for a rationale.) 
We will later see that this line also has physical significance. 

The general relation between the acceleration to steepness ratio $Y$ 
and the $w-w'$ track is 
\beq 
w'=3(1+w)\left[w+\frac{1-Y}{1+Y}\right]=3w(1+w)+3(1+w)\frac{1-Y}{1+Y}. 
\eeq 
Again we have a parabolic behavior.  The coasting line has $Y=0$, and 
the null line corresponds to $Y\to\infty$.  One could use either variable 
$X$ or $Y$ throughout the phase plane, since 
\beq 
X=-3\frac{Y}{1+Y}\quad ; \quad Y=-\frac {X}{X+3}, 
\eeq 
but this somewhat obscures the physics of friction and steepness.  (That 
said, we note that the thawing/freezing boundaries are fairly symmetric 
in $X$, with the outer boundaries at $X=\pm3/2$ and the inner ones at 
$X\approx\pm1/2$.)\footnote{Recall from Paper 1 that the upper bound on 
the freezing region is not sharply defined, and extends somewhat above 
the $w'=w(1+w)$ line shown for convenience in this paper.} 

Note that it is not legitimate to assume tracking behavior (where the 
equation of state is constant and related to the dominant component's 
equation of state) to impose limits on regions in the phase plane, 
as for example \cite{chiba,scherrer} did to try to tighten the 
constraints of Paper 1.  For one thing, not all freezing models 
need start as trackers.  Secondly, just because the most negative value 
of $w'$ lies above some boundary curve does not ensure that the entire 
trajectory remains above the curve.  
Most importantly, the tracking 
approximation breaks down as the dark energy first becomes significant, 
so it is inapplicable for much of the observable dynamical history.

\section{More Specific Dynamics \label{sec:more}} 

In addition to analyzing general behavior through the Klein-Gordon equation 
terms, we can investigate the properties of the phase plane or specific 
track families in terms of other variables.  These could include working 
from the equation of state $w-w'$ relation directly or the cosmic expansion 
history $a(t)$.   While not quite as insightful as the physics 
motivated driving and drag terms, they can highlight interesting properties.

\subsection{Mocker models \label{sec:mocker}} 

Consider a model with dynamics given by $w'=Cw(1+w)$.  Note that while 
such an equation forms the boundaries of the freezing region, freezing 
models do not follow such a trajectory but rather are almost orthogonal 
to such tracks (at least initially).  So we are talking about 
fundamentally different models.  The behavior of the dark energy 
equation of state and density follow 
\beqa 
w(a)&=&-1+\left[1-\frac{w_0}{1+w_0}a^C\right]^{-1} \\ 
\rho_{\rm de}(a)&=&\rho_{\rm de}(1)\,\left[(1+w_0)a^{-C}-w_0\right]^{3/C}, 
\eeqa 
where $w_0$ is the equation of state today. 

In the past, $a\ll1$, the component will act like additional nonrelativistic 
matter, with $w=0$, while in the future it will approach 
a cosmological constant.  Since such dark energy sometimes looks like 
dark matter and sometimes like a cosmological constant, we call this a 
``mocker'' model.  These are basically what are known as ``quartessence'' 
models (see \cite{makleroliveirawaga} for an overview), of which the 
Chaplygin gas is one 
example.  However, we develop them directly from the phase space dynamics 
rather than an ansatz for the pressure, so the dynamical behavior is 
more general.  For example, a 
constant pressure model could be a cosmological constant, or could be 
a mocker with $C=3$.  Note $C=3$ gives precisely the expression for the 
lower boundary of the freezing 
region (or is $Y=1$ or $X=-3/2$ in the notation of \S\ref{sec:kglines}). 
We name the $w'=3w(1+w)$ curve the constant pressure line (also see 
\cite{scherrer} in the context of barotropic fluids). 

Such combined behavior models are heir to all 
the usual problems of trying to unify dark matter and dark energy, 
e.g.\ growth instabilities of density perturbations 
\cite{sandvik,amendola0509099}.  Analysis of perturbations requires 
knowledge of the full theory, however. 
Merely from the phase plane dynamics, though, we can see trouble arising 
for such unified models. 
As $C$ gets smaller, the model moves along its trajectory more quickly, 
acting less like dark matter except at very early times, 
$1+z\gg[-w_0/(1+w_0)]^{1/C}$.  Conversely, 
as $C$ gets larger, acceleration of the cosmic expansion occurs later 
and the model becomes a poorer fit to a host of (purely geometric) 
observations such as supernova distances and the distance to the 
CMB last scattering surface.

\subsection{Relation to parameterized $w(a)$ \label{sec:parw}} 

The approach taken in this article is to examine dark energy dynamics 
directly in the phase plane $w-w'$, where a time variable runs along 
each trajectory.  It is useful to see the relation 
of standard parametrizations in terms of the temporal behavior, 
i.e.\ $w(a)$, to this approach.  

The standard two parameter function $w(a)=w_0+w_a(1-a)$ was shown by 
Linder \cite{linprl,linpr} to provide an excellent approximation to 
exact solutions of the Klein-Gordon equation in a wide variety of 
models.  In this ansatz we have $w'=-aw_a=w-w_\infty$, where the 
high redshift equation of state $w_\infty=w_0+w_a$.  
This describes a straight 
line of slope 1 in the $w-w'$ plane, and can be rewritten as 
as $w'=(1+w)-(1+w_\infty)$.  In particular, if $w_\infty=-1$ 
we have exactly the behavior of thawing models (lying along lower bound of 
that region).  From Fig.~2 of Paper 1, we see as well that many tracking 
models that fit present data (i.e.\ $\Omega_{\rm de}\sim0.7$ and $w<-0.8$) 
are reasonably well described, on average, by a line of slope unity.  Of 
course this approximation will break down in the future, as the 
field freezes more fully, turning toward the cosmological constant; 
at the same time this $w(a)$ ansatz loses validity as it moves toward 
ever more negative $w$.  However, since data only exist toward the past, 
we see why the $w_a$ parametrization is an excellent approximation. 

To retain boundedness for both the past and present, as well as to 
allow more dramatic dynamics (essentially slopes other than unity in 
$w-w'$), one could use the ``e-fold'' model of \cite{eospar} or ``kink'' 
model of \cite{corakink}.  Both utilize four parameters for their 
description.  The e-fold model has a more transparent translation to 
$w'=dw/d\ln a$ since it also uses dynamics in terms of $\ln a$: 
\beq 
w(a)=w_f+\frac{\Delta w}{1+(a/a_t)^\tau}, 
\eeq 
where $\tau$ is the transition rapidity in units of units of 
e-folds $\ln a$, $a_t$ is the transition scale factor, $w_f$ is the 
asymptotic future value, and $\Delta w=w_p-w_f$ is the difference 
between asymptotic past and future values. 

In the $w-w'$ phase plane we have 
\beq 
w'=-\tau(w-w_f)\left(1-\frac{w-w_f}{\Delta w}\right). 
\eeq 
Note that as we found in the Klein-Gordon equation analysis of 
\S\ref{sec:kglines} the equation for $w'$ is quadratic in $w$. 
We can identify several special cases.  If the asymptotic future 
state is deSitter ($w_f=-1$), then $w'=\tau(\Delta w^{-1}-1)(1+w) 
+(\tau/\Delta w)\,w(1+w)$.  This looks like the sum of a thawing 
model and a model in the freezing region, i.e.\ the dark energy can be 
viewed as the sum of two components.  If we further take $\Delta w=1$, 
then we remove the thawing component and end up with $w'=\tau\,w(1+w)$ 
-- a mocker model with $w_p=0$ and $w_f=-1$.  

Starting instead with 
an asymptotic past state of $w_p=-1$ gives $w'=\tau\,(1+w)(w-w_f)/(-1-w_f)$. 
In the limit $w_f\to\infty$ (i.e.\ not worrying about the region where 
there is no data) this gives a thawing model $w'=\tau(1+w)$.  Thus the 
four parameter e-fold ansatz is also quite versatile.  The rapidity 
parameter is directly related to both the slope of and the velocity 
along the phase space trajectory, and ties in with the steepness of 
the scalar field potential, as we saw in \S\ref{sec:kglines} with the 
PNGB models where the slope was proportional to the inverse of the 
symmetry scale $f$. 

Finally, one could invert the situation and go from a 
parametrization in the phase plane to 
derive the function $w(a)$.  For example, a track $w'=A(1+w)+B(1+w)^2$, 
which we have seen is a common form, implies 
\beq 
1+w=(1+w_0)/\left[(1+x)a^{-A}-x\right], 
\eeq 
where $x=(B/A)(1+w_0)$ defines the present along the trajectory 
(equivalently the dimensionless matter density $\om$ today).  Note 
that while the trajectory has two 
parameters, the equation of state has three parameters since we 
must have a parameter running along the track.  At high redshift, 
if $A>0$ then $w\to-1$ and we have a thawing model, asymptotically 
independent of $B$.  If $A<0$ then $w(z\gg1)=-1+(-A/B)$ and $w'\to0$, 
i.e.\ it begins like a tracking model.  It reaches a minimum 
$w'_{\rm min}=-A^2/(4B)$ at $w_\star=-1-A/(2B)=[-1+w(z\gg1)]/2$; 
that is, the trajectory is a parabola from its tracking value of the 
equation of state to its future, cosmological constant value.  A mocker 
model is the special case $C=-A=B$.  For 
completeness, we give the dark energy density: 
\beq 
\rho_{\rm de}(a)=\rho_{\rm de}(1)\,(1+x-xa^A)^{3/B}. 
\eeq 

Of particular interest is the ``leveling'' model where, in loose 
physical analogy to the inflationary power spectrum tilt $n-1$ being 
driven to zero by large numbers of e-folds of expansion, the 
equation of state tilt $1+w$ is driven to zero by the deSitter 
expansion as the energy density approaches a certain constant value, 
$\rho_f$.  That is, take $1+w=D[\rho_{\rm de}(a)-\rho_f]$.  This is 
equivalent to the above parabolic model with $A=-3D\rho_f$ and $B=-3$. 
Another interesting parabolic track is the coasting line $w'=3(1+w)^2$. 
This corresponds to not a leveling but a tilting, with 
$1+w=(1+w_0)\,[\rho_{\rm de}(1)/\rho_{\rm de}(a)]$, so $w$ is tilted 
away from $-1$ as the energy density decreases.

\subsection{Acceleration and jerk \label{sec:jerk}}

One could leave behind the physics of the accelerating phenomenon 
and instead use variables in terms of the acceleration itself, though 
this seems less appealing.  The deceleration parameter 
\beq 
q=-a\ddot a/\dot a^2=\frac{1}{2}+\frac{3}{2}w\Omega_{de}(a), 
\eeq 
and the jerk 
\beq 
j=a^2\, \dddot a/\dot a^3=1-\frac{3}{2}\Omega_{de}(a)\,[w'-3w(1+w)]. 
\eeq 
We also have $j=q+2q^2-q'$.  Note that interpreting $q$ and $j$ or $q'$ as 
Taylor expansions of the expansion is of strictly limited use (since 
observations span $\Delta\ln a\sim{\cal O}(1)$) and can be dangerous 
\cite{noqexp}. 

Furthermore, there is the same ambiguity there was with using pressure 
as a variable.  We note that any model where it touches the constant 
pressure line 
$w'=3w(1+w)$ has $j=1$; this is equivalent to $X=-3/2$ in the notation 
of \S\ref{sec:kglines}.  The two standard special cases of $j=1$ lie 
at the ends of this line: an Einstein-de Sitter universe with $w=0$ 
``dark energy'' and a $\Lambda$CDM universe with cosmological constant dark 
energy. 

The constant pressure line is also related to the adiabatic sound speed of 
the dark energy.  (Note this is not the true sound speed of perturbations 
arising from the microphysics of whatever the dark energy is.)  The 
adiabatic sound speed 
\beq 
c_a^2=\frac{\dot p}{\dot\rho}=w-\frac{1}{3}\frac{w'}{1+w}, \label{eq:ca} 
\eeq 
and we see it vanishes for $w'=3w(1+w)$.  On the null line, the adiabatic 
sound speed equals the speed of light (the same as the true sound speed 
for a canonical scalar field).  Models below the null line would need 
to have $c_a^2>1$.

\section{Beyond Scalar Fields \label{sec:beyond}} 

For the cosmic expansion dynamics we can always define an 
effective equation of state even if the accelerating mechanism is 
not a scalar field \cite{linjen}, through 
\beq 
w_{\rm eff}=-1-\frac{1}{3}\frac{d\ln\delta H^2}{d\ln a}, \label{eq:weff}
\eeq 
where $\delta H^2=(H/H_0)^2-\om a^{-3}$ is the unknown part of the 
Hubble parameter, that not due to matter.  So it is of interest to 
see to what extent the dynamical behaviors we have discussed carry 
over to the $w_{\rm eff}-w'_{\rm eff}$ plane.  That is, are freezing 
and thawing behaviors more general than for scalar fields, and do 
the null and coasting lines still play a role? 

Due to the diversity of possible accelerating physics we do not 
present a general analysis of these important questions 
but rather calculate some specific cases. 

\subsection{Scalar-tensor gravity \label{sec:scatens}}

Scalar-tensor theories modify the Einstein-Hilbert action with both an 
additional scalar field and a coupling to Ricci scalar curvature $R$. 
These are of great interest as a comparison in tests of general 
relativity and also because gravitational theories involving 
a nontrivial function of the scalar curvature can be transformed 
to scalar-tensor theories.  See \cite{scatensreview} for a general 
introduction. 

We consider coupling of a general form in the scalar field, but linear 
in the curvature.  So the general relativistic $R/(8\pi G)\to F(\phi)R$.  
One then 
obtains the usual Friedmann expansion equations, with extra terms 
giving an effective dark energy density \cite{bmp} 
\beq 
\rho_{\rm ST}=V(\phi)+(1/2)H^2(q-1)(q+5)F_\phi^2+3H^2[(8\pi G)^{-1}-F], 
\label{eq:rhost} 
\eeq 
where $q$ is the deceleration parameter, $F_\phi=dF/d\phi$, and the last 
term involves the change of the gravitational strength from Newton's 
constant $G$. 

From this density one can then define $\delta H^2=8\pi G\rho_{\rm ST}/(3H^2)$ 
(note that we use the usual $G$ here since the deviation is absorbed into 
$\rho_{\rm ST}$ as mentioned above).  From this the equation of state 
$w_{\rm eff}$ and its variation $w'_{\rm eff}$ can be calculated.  A key 
quantity will be $F/F_\phi^2\equiv\omega_{JBD}$.  This is the 
Jordan-Brans-Dicke parameter and its inverse must be very small 
according to solar system tests.  Expanding $F$ about the present, 
\beqa 
F(a)&\approx&(8\pi G)^{-1}-(1-a)F_\phi \dot\phi/(aH) \\ 
&\approx&(8\pi G)^{-1}-z(1-q)F_\phi^2. 
\eeqa 
So the ratio of the second to the first term ($\sim\omega_{JBD}^{-1}$) 
is small, and gravity is nearly Einsteinian. 

But this means that the first term in $\rho_{\rm ST}$ dominates (unless 
$8\pi GV/H^2\ll1$, but then it doesn't affect the expansion and there is 
no acceleration).  Thus, the restriction of the scalar-tensor theory by 
solar system constraints means that its effective equation of state must 
be very close to a cosmological constant -- within $\sim\omega_{JBD}^{-1}$. 
Since $\omega_{JBD}^{-1}<2.5\times 10^{-5}$, this would be rather 
challenging to distinguish from a cosmological constant with cosmological 
observations!  One possible loophole is if the solar system limits on 
the scalar coupling should not be applied to a cosmological situation 
because of the different spacetime backgrounds with very different 
scalar curvatures.  This arises for example in chameleon scenarios 
\cite{chameleon}.  The most stringent cosmological bounds on varying $G$ 
arise from primordial nucleosynthesis and give 
$\omega_{JBD}^{-1}\lesssim 3\times10^{-3}$ \cite{gvarycos} 
(but see \cite{nsvary}). 

Calculation of the effective phase plane parameters finds 
\cite{bmplinder} 
\beqa 
w_{\rm eff}(z=0)&=&-1+0.46/\omega_{JBD} \\ 
w'_{\rm eff}(z=0)&=&-0.36/\omega_{JBD}. 
\eeqa 
While even with only the cosmological bounds on $\omega_{JBD}$ these 
are quite close to the cosmological constant in the phase plane, 
it is interesting to note that the current values lie along 
$w'=0.78w(1+w)$, in the freezing region.  Indeed its 
trajectory is a freezing one, with scalar-tensor theories asymptotically 
attracted to general relativity \cite{damourst,rboost}, and to $w=-1$. 
One last thing to note, however, is that because scalar-tensor theories 
possess anisotropic stress, the growth of density perturbations will 
be modified from the quintessence case (see, e.g., \cite{stgrowth} and 
references therein).

\subsection{Braneworld cosmology and $H^\alpha$ \label{sec:brane}} 

\begin{figure}[!hbt]
\begin{center}
\psfig{file=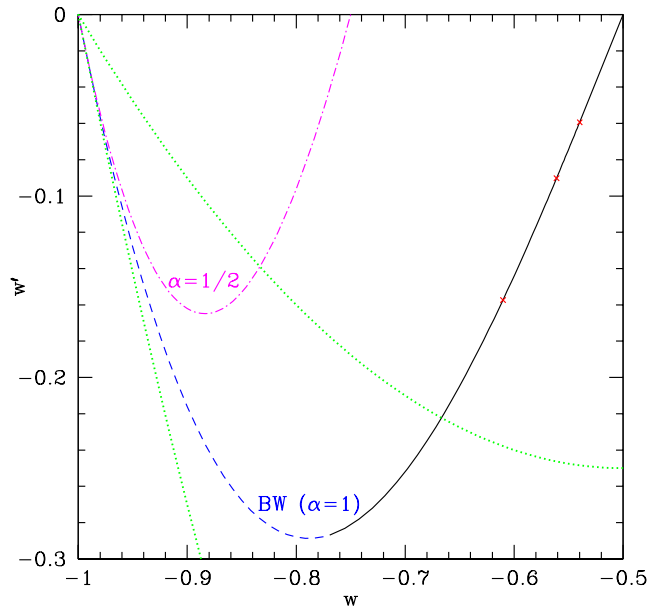,width=3.4in}
\caption{The braneworld model of modified gravity behaves like a 
freezing scalar field.  The solid black line shows its trajectory: 
starting at high redshift at $w=-1/2$, $w'=0$, with the red crosses 
indicating $z=3$, 2, 1 before the line ends at the present.  The dashed 
blue line shows its future behavior, approaching a cosmological constant, 
and lying within the quintessence freezing bounds of $w'=3w(1+w)$ and 
$w'=w(1+w)$ given by the dotted green curves.  
A model with a generalized, extra $H^\alpha$ term (for the 
braneworld, $\alpha=1$) in the Friedmann expansion equation 
looks similar; the magenta, dot-dash curve is for $\alpha=0.5$. 
}
\label{fig:bw}
\end{center}
\end{figure}

In a braneworld cosmology \cite{dgp,deffayet}, effective acceleration 
appears due to a weakening of gravity on large scales as it ``leaks'' 
from our brane into a higher dimensional bulk.  The Friedmann expansion 
equation becomes 
\beq 
H^2-H/r_c=8\pi G\rho_m/3, 
\eeq 
where $r_c$ is the crossover distance and $\rho_m$ the matter density. 
The effective equation of state due to the modification is $w_{\rm eff}= 
-[1+\om(a)]^{-1}$ \cite{lueeos}.  Its trajectory in the phase plane is 
plotted in Fig.~\ref{fig:bw}.  Note that it looks like a freezing model, 
and will indeed approach a cosmological constant in the asymptotic future. 

The position along the 
trajectory is a time variable, so taking the present to be, say, $\om=0.2$ 
would extend the solid curve in Fig.~\ref{fig:bw} slightly further 
(since the figure uses $\om=0.3$). 
We also see why it is so well approximated by a $w_0-w_a$ model, as 
discussed in \S\ref{sec:parw}.  (Recall, however, that $w_a$ is actually 
defined at $z=1$, not $z=0$, to give the best physics fit \cite{linprl}.)

We can further generalize the modification to $\delta H^2\sim H^\alpha$ 
\cite{dvaliturner}.  Then we find 
\beqa 
w_{\rm eff}&=&-\left[1-\frac{\alpha}{\alpha-2}\om(a)\right]^{-1} \\ 
w'_{\rm eff}&=&3w(1+w)[1-(2/\alpha)(1+w)]. \label{eq:wpalpha} 
\eeqa 
Recall that the braneworld case above corresponds to $\alpha=1$. 
For acceleration today (with $\om=0.3$), we require $\alpha<1.57$; 
for $w<-0.8$ today we require $\alpha<0.91$.  Note that all $H^\alpha$ 
modified gravity models will look similar (one does require $\alpha<2$ 
for a negative equation of state at early times).  See \S\ref{sec:track} 
for discussion 
of their tracking behavior.  In particular, they all approach 
the cosmological constant along $w'=3w(1+w)$, what was called the 
constant pressure line for scalar fields.  Their tracks must always 
lie between $w'=3w(1+w)$ and the $w'=0$ axis.  When $\alpha<0$, the 
trajectories switch to the phantom regime with $w<-1$, but the bounds 
still hold.

\section{Polytropic Dark Energy \label{sec:poly}} 

An interesting, if phenomenological, way of obtaining acceleration 
is to modify the Friedmann expansion equation but keeping a pure 
matter universe.  While this leaves open important questions about 
its relation to fundamental theory and the growth of density perturbations, 
we can investigate some general aspects of the effective equation of state 
dynamics. 

Consider general functions of the matter density (sometimes referred 
to as barotropic models \cite{scherrer}) 
\beq 
H^2=(8\pi G/3)\,g(\rho)=(8\pi G/3)\,[\rho+f(\rho)]. 
\eeq 
The quantity $f(\rho)$ will act like an effective dark energy.  Using 
Eq.~(\ref{eq:weff}) we can define 
\beqa 
w_{\rm eff}&=&-1+\frac{d\ln f}{d\ln\rho} \\ 
w'_{\rm eff}&=&-3\,\frac{d^2\ln f}{d\ln\rho^2}, 
\eeqa 
and identical relations hold for the total equation of state $w_{\rm tot}$ 
and its variation $w'_{\rm tot}$ upon substituting $g$ for $f$. 

The simplest example of such a modification is $f\sim\rho^n$, the 
Cardassian model of \cite{freeselewis}.  From the above equations we 
see that it corresponds to a constant equation of state $w=-1+n$.  If 
we require $w<-0.9$ (as observations favor for a constant equation of 
state), then $n<0.1$; unfortunately $\rho^{1/10}$ does not obviously 
appear to be a natural modification of the Friedmann equation.  

We can investigate the phase space dynamics by relating $w'$ to $w$: 
\beq 
w'=3w(1+w)-3\frac{\rho^2}{f}\frac{d^2f}{d\rho^2}. 
\eeq 
Immediately we see that whether the effective dark energy lies below 
the freezing region or not depends on the sign of $d^2f/d\rho^2$. 
An equivalent question is whether the effective 
pressure is decreasing or increasing with time (since the $w'=3w(1+w)$ 
line is that of constant pressure).  

The model will follow the mocker model track $w'=3w(1+w)$ if $f=A+B\rho$. 
This form is equivalent to a redefinition of $\om$, e.g. $\om\to\om(1+B)$, 
plus a cosmological constant $A$.  As such it has a nonzero minimum in 
its effective potential, allowing it to reach the freezing boundary.  

It is important to remember that the analysis in this paper and Paper 1 
applies to the dynamics of the dark energy itself.  Trajectories in the 
$w_{\rm tot}-w'_{\rm tot}$ plane convolve the matter and dark energy 
components, mixing the dynamics and so not giving rise to the clear 
differentiations and regions found.  This is why Ref.~\cite{scherrer} appears 
to find a violation of the freezing bound and even null bound for some 
barotropic models; those are actually phantom models in the dark energy 
phase space, but are dragged by the matter to $w_{\rm tot}>-1$.  We 
make this more explicit later in this section. 

For a richer dynamical behavior we propose a class of modifications of 
the Friedmann expansion equation we call polytropic models \cite{freese02}.  
Here 
\beq 
H^2/(8\pi G/3)=g(\rho)=\rho\,[1+(\rho/\rhos)^{-\alpha}]^\beta. 
\label{eq:polyg}
\eeq 
At densities much greater than some crossover value $\rhos$, e.g.\ at 
high redshift, the Friedmann equation is standard.  At low densities, 
the expansion is modified, with $w_{\rm tot}$ asymptotically approaching 
$-\alpha\beta$.  If we want a future deSitter state, we could choose 
$\beta=1/\alpha$.  For just the effective dark energy equation of state, 
the value in the past is $w_{\rm eff}=-\alpha$, and in the future of 
course it dominates so $w_{\rm eff}=-\alpha\beta$. 

\begin{figure}[!hbt]
\begin{center}
\psfig{file=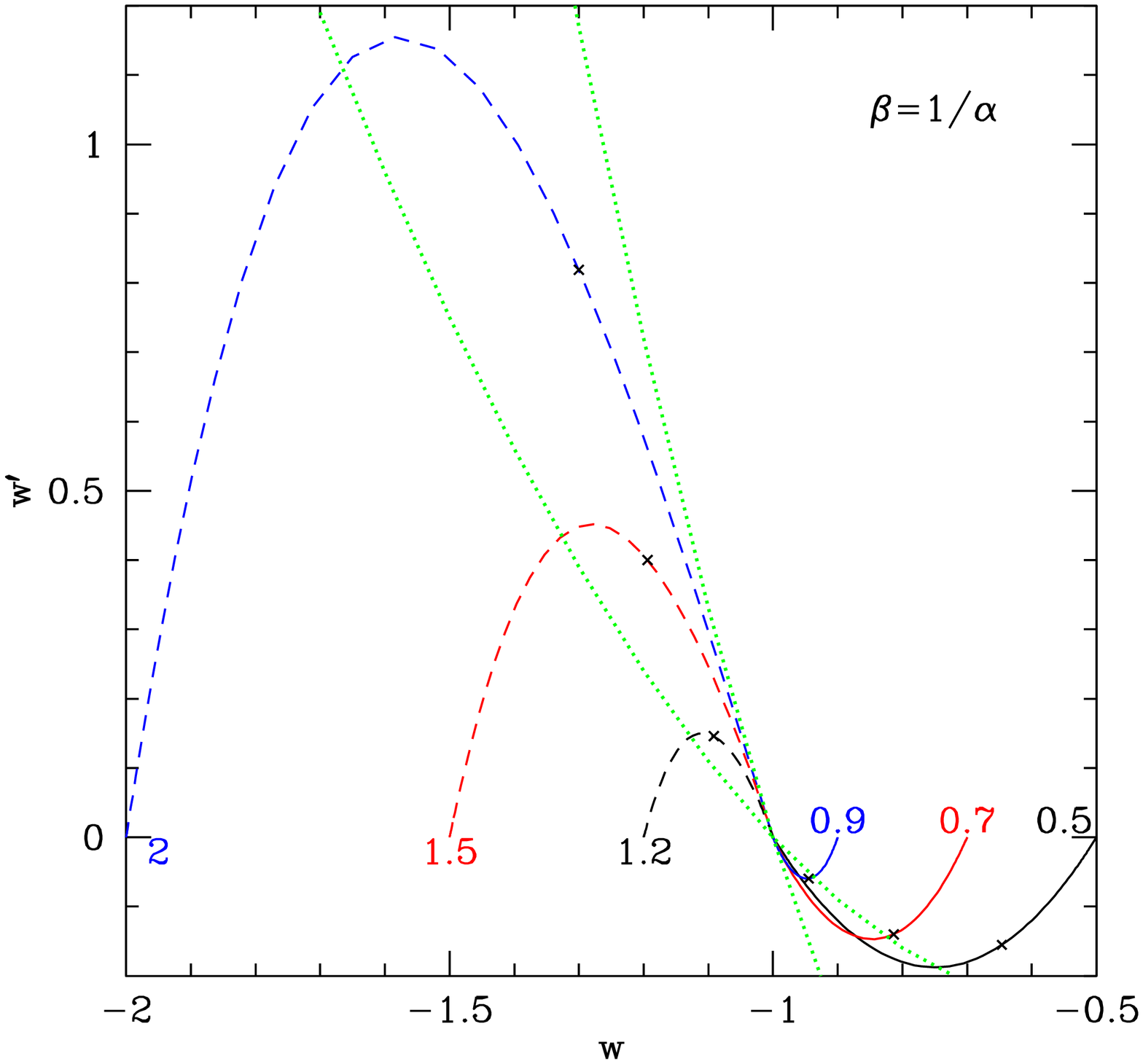,width=3.4in}
\psfig{file=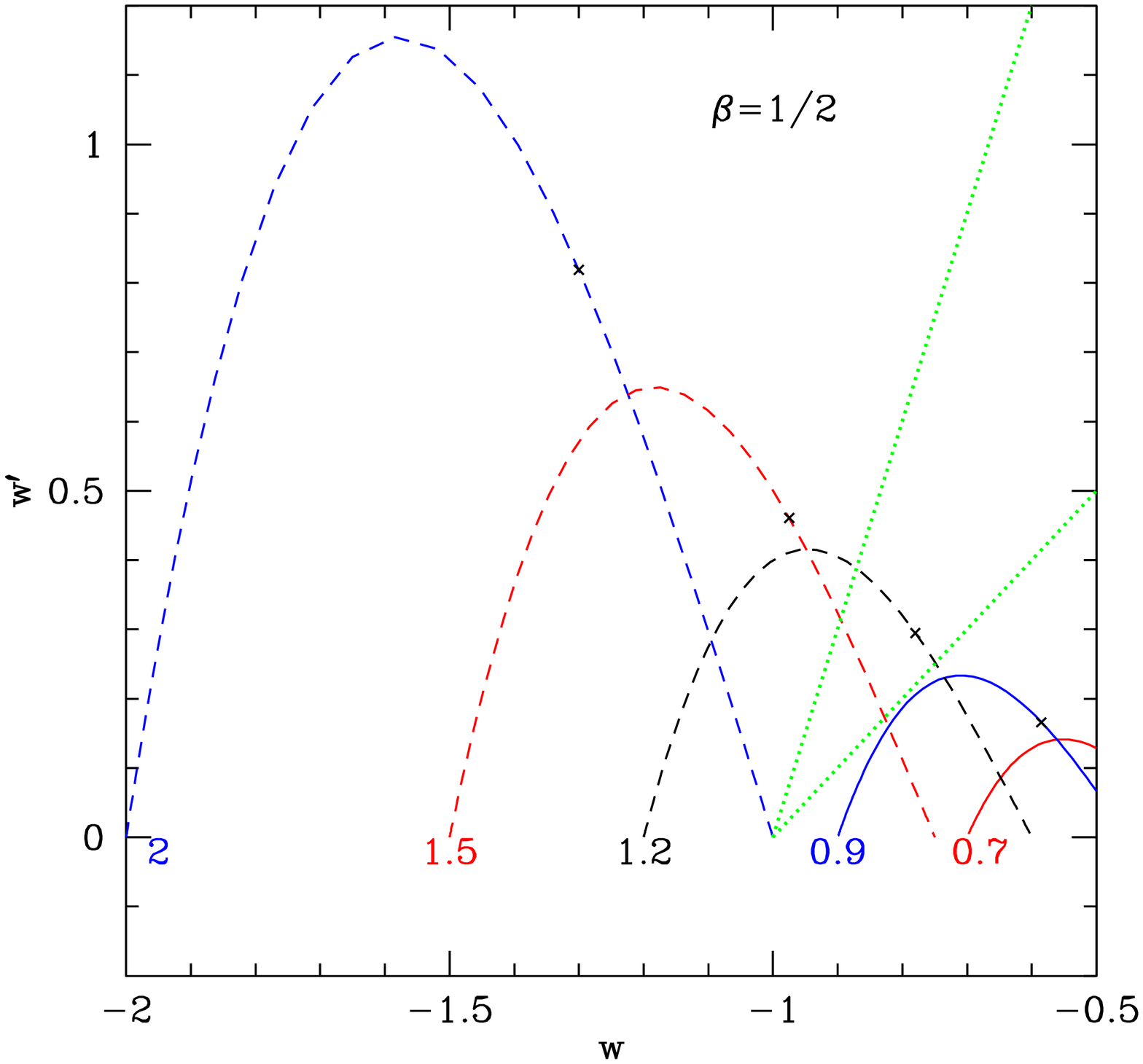,width=3.4in}
\caption{Polytropic models, with $\beta=1/\alpha$ in the first panel and 
with $\beta=1/2$ in the second panel.  Curves are 
labeled with $\alpha$ and crosses show the present.  Dotted curves show 
the canonical scalar field freezing region (first panel) or thawing 
region (second panel).  In the first panel, models evolve toward a 
cosmological constant, within the freezing region.  In the second panel, 
the asymptotic state is not deSitter (except for the $\alpha=2$ curve, 
which is the same in both panels) and the dynamics is distinct from a 
canonical scalar field. 
}
\label{fig:polyw}
\end{center}
\end{figure}

Figure~\ref{fig:polyw} illustrates the phase plane dynamics.  The first 
panel takes $\beta=1/\alpha$, so that the future state is deSitter. 
Models with $\alpha<1$ have $w\ge-1$ and act like freezing models, 
starting from a constant $w=-\alpha$ and today (marked by crosses) 
lying in the freezing region, before heading toward the cosmological 
constant.   Phantom models have $\alpha>1$ and act like mirror images 
of freezing models, even to lying within the phantom freezing region today.

In the second panel, we fix $\beta=1/2$.  While the models start at 
the same phase space point as the previous models with the same $\alpha$, 
now their endpoints are different.  Indeed for $\alpha<2/3$ the 
acceleration of the expansion is a temporary phase.  Furthermore, the 
trajectories with $\alpha\lesssim1.5$ do not lie in 
the freezing regime and both regular and phantom models have 
$w'>0$.  Such polytropic models without a deSitter future will be clearly 
distinguishable from both freezing and thawing quintessence. 

Can we give some physics motivation for the polytropic form, aside from 
its simplicity and proper asymptotic behavior?  When $\beta=0$, there is 
no modification; when $\beta=1$ we have the power law modification of 
the Cardassian case, with $n=1-\alpha$, hence a constant $w=-\alpha$. 
There are some motivations for power law modification from higher dimension 
theories (for $n<1$ see \cite{freeselewis}, the nonaccelerating $n=2$ 
arises in Randall-Sundrum brane scenarios \cite{randallsundrum}). 
When $\beta=1/2$, the modification is similar to that from a Chaplygin 
gas \cite{chaplygingas}, as we see below; this has claimed motivation 
from Born-Infeld actions and brane solutions \cite{bento}.  So at 
least the polytropic form unifies different prescriptions for dark energy. 

Note that as $\beta$ increases from zero, for fixed $\alpha$, the size of 
the ``hump'' in the trajectory decreases and the future value of $w$ 
moves back toward the initial value.  At $\beta=1$ the trajectory 
collapses to a point at $w=-\alpha$, $w'=0$.  For even larger $\beta$, 
the hump is flipped (i.e.\ the sign of $w'$ changes) and again 
increases in size, with the future value of $w$ drawing away to more 
negative values. 

\begin{figure}[!hbt]
\begin{center}
\psfig{file=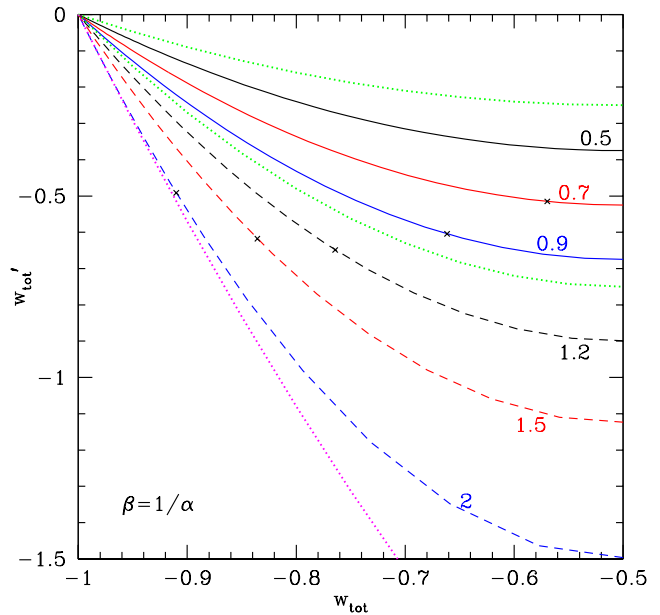,width=3.4in}
\caption{Polytropic models as Fig.~\ref{fig:polyw}a, with $\beta=1/\alpha$, 
but now viewed in the total equation of state phase plane.  Curves are 
labeled with $\alpha$ and crosses show the present.  Dotted curves show 
the {\it dark energy\/} freezing region bounds (light green) 
and the null line (dark magenta).  Note that those 
models phantom in the dark energy here appear with $w_{\rm tot}>-1$ but 
lie below $w'_{\rm tot}=3w_{\rm tot}(1+w_{\rm tot})$.  One should {\it not\/} 
apply the dark energy dynamics constraints to the total equation of state 
plane. 
}
\label{fig:polywtot}
\end{center}
\end{figure}

When we plot the same models as in the first panel of Fig.~\ref{fig:polyw}, 
but in the $w_{\rm tot}-w'_{\rm tot}$ plane, in Fig.~\ref{fig:polywtot}, 
we see that the models 
that were phantom in the effective dark energy lie in the region 
$w'_{\rm tot}<3w_{\rm tot}(1+w_{\rm tot})$.  Furthermore, when $\alpha>2$ 
they can even lie below what was the null line $w'=-3(1-w^2)$.  In a 
nice analysis of barotropic models, Scherrer \cite{scherrer} noted 
something similar (his barotropic models are a function of an arbitrary 
component density not necessarily matter density).  
For a barotropic perfect fluid the adiabatic sound speed is the physically 
relevant sound speed for perturbations (but not in the quintessence case, 
or in a general multicomponent case), and Scherrer's bound of $w'<3w(1+w)$ 
holds for $c_a^2>0$ (cf.\ Eq.~\ref{eq:ca} here).  In general, however, 
this is not a violation of the bounds of this article and Paper 1 
because it occurs only when the adiabatic assumption holds, e.g.\ when 
viewing the total equation of state, not the 
properties of a (non-adiabatic) effective dark energy.

Suppose, however, we chose to fit the dark energy component itself, 
rather than full energy density entering the Friedmann equation, 
by the polytropic form Eq.~(\ref{eq:polyg}).  This is somewhat strange 
to do, since then the effective dark energy contains a matter-like part, 
in addition to the pure matter density, and the polytrope was designed 
to be the modified Friedmann equation as a whole.  If we do so, though, 
then the dark energy equation of state phase plane (and not the total 
equation of state) is represented by the curves in Fig.~\ref{fig:polywtot}. 
Moreover, the curve with $\alpha=2$, $\beta=1/2$ is the trajectory of 
the Chaplygin gas.  The generalized Chaplygin gas with pressure 
$p\sim-\rho^{-\alpha_{gcg}}$ corresponds to polytropic dark energy 
with $\alpha=\alpha_{gcg}+1$, $\beta=1/\alpha$.  Whenever $\alpha\beta=1$ 
(cf.\ \cite{gondolofreese}) 
we have mocker behavior with $w'=3\alpha w(1+w)$.   As stated above, 
however, taking the dark energy itself to be polytropic means hiding both 
a cosmological constant (if $\beta=1/\alpha$)  and a spurious matter 
density within the dark sector.

\section{Hierarchy parameters} \label{sec:slow} 

Returning to more general dynamics, we saw in \S\ref{sec:scadyn} and 
in particular from Fig.~\ref{fig:wwpscale} 
that we do not have the standard inflation slow roll perturbative expansion 
parameter.  Explicitly, we do not have small 
$\vp/V$ in the freezing or thawing regions unless $w\lesssim-0.995$; 
more generally, $-M_P\vp/V\gtrsim 5\sqrt{1+w}$ today. 
We examine here whether we can substitute a physics based hierarchy 
of dynamics parameters. 

\subsection{Slope parameters \label{sec:slope}}

The Klein-Gordon equation can be rewritten dimensionlessly as 
\beq 
\phi''+(2-q)\phi'=-\vp/H^2\equiv \eta_1,
\eeq 
where as before a prime denotes a derivative with respect to $\ln a$, 
and $q=-a\ddot a/\dot a^2$ is the deceleration parameter.  In terms of 
the $w-w'$ dynamics equation, 
\beqa 
w'&=&-3(1-w^2)-\sqrt{2(1-w^2)} \,\vp/(HV^{1/2})\\ 
&\equiv& -3(1-w^2)+\sqrt{2(1-w^2)} \,\eta_2. 
\eeqa  

So the parameters $\eta_1$, $\eta_2$ are called out by the physics. 
If they are small, one could solve the equations perturbatively.  
Note that $\eta_2=\eta_1(V/H^2)^{-1/2}$ so we always have 
$\eta_2>\eta_1$.  We violate the lower bound of the freezing region, 
$w'<3w(1+w)$, when 
\beq 
\eta_2^2<\frac{9}{2}\frac{\epsilon}{2-\epsilon}, 
\eeq 
where $\eps=1+w$ is the tilt parameter. 
The analogous condition such that $w'<0$ is $\eta_2^2<(9/2)\eps(2-\eps)$, 
and such that field is decelerating ($\ddot\phi<0$) is 
$\eta_2^2<18\eps/(2-\eps)$.  

Neither $\eta_1$ nor $\eta_2$ are particularly small unless $\eps=1+w$ is. 
For example $\eta_2\gtrsim 1.5\sqrt{\eps}$, giving $\eta_2>0.1$ for 
$w>-0.995$.  Even $\eta_1>0.1$ for $w>-0.94$.  The hierarchy among 
the parameters is fixed for the region of interest: $\eps<\eta_1<\eta_2$, 
so there is no phase space classification in this respect, as there is 
for large field, small field, and hybrid models in inflation \cite{kinney}. 
However that hierarchy involved the second derivative of the potential, 
so it is worth a brief look at that quantity. 

\subsection{Tracking parameter \label{sec:track}}

The tracking parameter is defined to be 
\beq 
\Gamma\equiv\frac{V\vpp}{\vp^2}. 
\eeq 
This is not generally a small parameter either.  Indeed, 
models whose energy density tracks \cite{tracker} the evolution of 
the dominant energy component fulfill the conditions 
\beq 
\Gamma>1\qquad;\qquad \frac{d\ln(\Gamma-1)}{d\ln a}\ll 1. 
\eeq 
Within the class of tracking models (so now a particular subset of scalar 
field cosmologies), at high redshifts within the matter dominated epoch 
the field obeys 
\beq 
\Gamma=1-\frac{w}{2(1+w)}. \label{eq:gammaw}
\eeq 
(Note the equation of state deviation $1+w$ at high redshift may not be 
small.) 

In \S\ref{sec:brane} we found that $H^\alpha$ models (including 
braneworlds) follow freezing trajectories.  This is not surprising 
because they are basically trackers.  A component starting with 
constant $w$ at early times is equivalent there to a modification 
$H^\alpha$ with $\alpha=2(1+w)$.  The tracking parameter 
$\Gamma=(2+\alpha)/(2\alpha)$ and it initially acts like an inverse 
power law potential $V\sim\phi^{-n}$ with $n=2\alpha/(2-\alpha)$. 

To relate the dynamics of the time variation to the tracking condition, 
we invert Eq.~(\ref{eq:gammaw}) to write 
\beqa 
w&=&-2(\Gamma-1)/[1+2(\Gamma-1)] \\ 
w'&=&\frac{dw}{d\ln a}=w(1+w)\frac{d\ln(\Gamma-1)}{d\ln a}. 
\eeqa 
However, this is of limited use since we are unlikely to be able to 
probe $w'$ in the high redshift, $z\gg1$, regime where tracking might hold.  
Strong acceleration today, with $w\lesssim-0.7$, requires the breakdown of 
tracking. 

However, the analogy to $H^\alpha$ models presents an important insight into 
why the general freezing region is bounded below by $w'=3w(1+w)$.  At early 
times the contribution to the Friedmann expansion equation by the dark 
energy is small and $\delta H^2\sim a^{-3(1+w)}\sim H^{2(1+w)}$.  That is, 
the effective $\alpha\approx 2(1+w)$.  One can generalize this to 
$\alpha=2\langle 1+w\rangle$ when time variation of the equation of 
state becomes relevant, where angle brackets denote an averaging over 
$\ln a$.  At late times, the dark energy density dominates the expansion, 
$\delta H^2\sim H^2$, as it approaches $w=-1$ (freezes). 
For any epoch we can define an instantaneous 
value of $\alpha$.  Equation~(\ref{eq:wpalpha}) then gives the relation 
for $w'$.  As freezing models approach $w=-1$, Eq.~(\ref{eq:wpalpha}) 
indicates they should do so along $w'=3w(1+w)$.  Furthermore, since the 
bracketed term is less than one, then this trajectory represents a 
general lower 
bound to the freezing region.\footnote{One caveat involves dark energy 
models that possess an internal cosmological constant, i.e.\ nonzero 
minimum to the potential, or otherwise act as the sum of two or more 
components.  These cannot be represented as $H^\alpha$ models and the 
freezing bound does not apply.}


\subsection{Dynamics, Mass, and Spatial Inhomogeneities} \label{sec:inhom} 

Dynamical models must also possess 
spatial inhomogeneities in the field at some level.  The equation for 
these is given by perturbation of the Klein-Gordon equation (\ref{eq:kg}), 
\beq 
\delta\ddot\phi+3H\delta\dot\phi+(k^2+\vpp)\delta\phi=-\dot h\dot\phi/2, 
\label{eq:pert}
\eeq 
where $k$ is the wavenumber and $h$ is the trace of the metric 
perturbation \cite{ma}.  Just as matter density perturbations are damped 
on scales below the Jeans length related to the sound speed in the 
background medium, so the spatial inhomogeneities in the scalar field 
will be absent on length scales less than that corresponding to the 
effective mass $\sqrt{\vpp}$. 

Using eq.\ (\ref{eq:wp}) and $\vpp=\dot \vp/\dot\phi$ we can calculate 
the critical mass scale (also see \cite{caldwellvpp}), with 
\beqa 
\vpp/H^2 &=& (2+3w+2q)\frac{w'}{1+w}+
\frac{1}{4}\left(\frac{w'}{1+w}\right)^2 \nonumber \\ 
&\,& -\frac{1}{2}\frac{w''}{1+w}+\frac{3}{4}(1-w)(5+3w+2q). 
\eeqa 
Note $w'/(1+w)$ and $w''/(1+w)$ are well behaved and generally nonzero 
as $w\to-1$. 

This shows that the goal of exploring the temporal and spatial dynamics 
of dark energy runs into double jeopardy.  If the time variation is 
weak, $|w'/(1+w)|\ll1$, then the effective mass $m=\sqrt{\vpp}\lesssim H$. 
(The scale $H$ today corresponds to $10^{-33}$ eV; dark energy would be a 
very light scalar field).  
This means the Compton wavelength of the scalar field perturbations 
is larger than the horizon, and so spatial inhomogeneities are also 
difficult to detect.  Thus, for $|w'|<1+w$, i.e.\ between the thawing and 
freezing regions there is a ``dead zone'' of phase space, 
where we can detect neither time variation nor spatial inhomogeneity. 

For appreciable time variation, $|w'|>1+w$, one can have $m>H$ and 
so the possibility of subhorizon clustering.  However for models within 
the freezing or thawing regions, one is restricted to $m\lesssim 2H$ 
so this could only occur on the largest scales (largest angles or 
lowest multipoles). 
Note that as the field approaches $w=-1$, the mass stays nonzero 
(except it vanishes along the upper boundary of thawing and along the 
null line).  However, the amplitude of spatial perturbations vanishes 
as can be seen from Eq.~(\ref{eq:pert}) with $\dot\phi=0$. 
These properties make scalar field inhomogeneity an extremely begrudging 
probe of the nature of dark energy, much less friendly than the dynamics.

\section{Conclusion} \label{sec.concl} 

We have deepened and elaborated the understanding of the role that 
the dark energy dynamics, through the $w-w'$ phase plane, can play 
in leading our understanding of the nature of dark energy.  This 
includes the foundations of the null line, coasting line, 
constant pressure line, and phantom line dividing the phase plane 
into distinct, physical regions.  We also elucidate the uppermost 
and lowermost boundaries of the thawing and freezing regions. 

The physical structure has been extended beyond canonical scalar 
fields, including specific instances of modified gravity scenarios 
such as scalar-tensor, braneworld, and $H^\alpha$ models, and 
barotropic and polytropic generalizations of the Friedmann equation. 
We outlined similarities and differences with the scalar field case, 
showing that many act as freezing fields, and that we should be able 
to clearly distinguish certain models that do not possess a deSitter 
future.  Mocker models, implementing a unification of dark matter 
and dark energy, were shown to have difficulties purely from dynamical 
considerations, in addition to their problems in structure formation. 

Dark energy is demonstrated to be generically not amenable to a slow 
roll description -- a major difference from early universe inflation -- 
as one of its ``Goldilocks'' conundra.  This makes the dark energy 
problem in some sense even more challenging than the early universe. 
However, it also opens the possibility that if some physical bound 
can be placed on the flatness of the potential, e.g.\ due to quantum 
corrections, then this implies a barrier around the cosmological 
constant $\Lambda$ model.  This would offer hope, possibly accessible to next 
generation experiments, that dark energy could definitely be distinguished 
from $\Lambda$, if it is not $\Lambda$.  That would be exciting! 

The dynamics of the dark energy, in the form of the equation of state 
ratio $w$ and its time variation $w'$, provides powerful insight into 
the new physics behind cosmic acceleration.   Spatial inhomogeneities 
in the dark energy are seen to be much weaker and less forthcoming, 
unless one entertains direct couplings.  While we are not guaranteed 
to zero in on the physics -- there is a dead zone of minimal 
dynamics and possibly a ``confusion'' zone near the cosmological constant 
-- any highly precise and accurate result would be an enormous success 
in enlightening us on the dark universe. 

$\,$ \\ 

\section*{Acknowledgments} 

I benefited greatly from numerous discussions with my collaborator 
Robert Caldwell.  I also thank Carlo Baccigalupi, Robert Scherrer, and 
participants in the workshop Cosmological Frontiers in Fundamental 
Physics at Perimeter Institute (which I thank for hospitality), the 
workshop Dark Energy from Fundamentals (DarkFun 3, hosted by the SNAP 
Collaboration), and the LBNL particle theory group. 
This work has been supported in part by the Director, Office of Science,
Department of Energy under grant DE-AC02-05CH11231. 

\end{document}